\documentclass[10pt]{article}
\usepackage{todonotes}
\usepackage{subfigure}
\usepackage{amsmath}
\usepackage{amssymb}
\usepackage{cite}
\usepackage{hyperref}
\usepackage{lineno}
\usepackage{microtype}
\usepackage{indentfirst}
\DisableLigatures[f]{encoding = *, family = * }
\usepackage[labelfont=bf,labelsep=period,justification=raggedright]{caption}
\makeatletter
\renewcommand{\@biblabel}[1]{\quad#1.}
\makeatother

\def \rr {$R_{t}\:$}

\date{}

\begin{document}

% Title must be 150 characters or less
\begin{flushleft}
{\Large
\textbf{Estimating the Attack Ratio of Dengue Epidemics under Time-varying 
Force of Infection using Aggregated Notification Data}
}
\\
Flavio Code\c{c}o Coelho$^{1\ast}$, 
Luiz Max de Carvalho$^{2}$, 
\\
\bf{1} Escola de Matem\'atica Aplicada , Funda\c{c}\~ao Getulio Vargas (FGV), 
Rio de Janeiro -- RJ, Brazil.
\\
\bf{2} Programa de Computa\c{c}\~ao Cient\'ifica (PROCC), Funda\c{c}\~ao Oswaldo 
Cruz, Rio de Janeiro -- RJ, Brazil.
\\
$\ast$ E-mail: fccoelho@fgv.br
\end{flushleft}

\section*{Abstract}

Quantifying the attack ratio of disease is key to epidemiological inference and 
Public Health planning.
For multi-serotype pathogens, however, different levels of 
serotype-specific immunity make it difficult to assess the population at risk. 
In this paper we propose a Bayesian method for estimation of the attack ratio 
of an epidemic and the initial fraction of susceptibles using aggregated 
incidence data. 
We derive the probability distribution of the effective reproductive number, 
\rr, and use MCMC to obtain posterior 
distributions of the parameters of a single-strain SIR transmission model with 
time-varying force of infection.
Our method is showcased in a data set consisting of 18 years of dengue 
incidence in the city of Rio de Janeiro, Brazil.
We demonstrate that it is possible to learn about the initial fraction of 
susceptibles and the attack ratio even in the absence of serotype specific data.
On the other hand, the information provided by this approach is limited, 
stressing the need for detailed serological surveys to characterise the 
distribution of serotype-specific immunity in the population.

\section*{Introduction}

Dengue is an arthropod-borne febrile disease caused by a flavivirus with four 
serotypes which causes an estimated  $50$ million infections each 
year~\cite{Guzman2010}.
In humans, immunity against  a particular serotype is  considered permanent 
after the exposure and cross immunity to there serotypes is considered short 
lived~\cite{halstead_dengue_2007}.
As a consequence, the proportions of viral serotypes co-circulating at any 
point in time are strongly dependent on previous incidence patterns of the 
disease, which determine the number of individuals susceptible to each serotype 
at any point in time.
% LM: Maybe a citation here?

Dengue transmission is also modulated by environmental conditions, among which, 
temperature, due to its effects on the 
vector reproduction, stands out as a strong predictor of 
incidence~\cite{honorio_temporal_2009,wu_higher_2009}.
In places with sufficient seasonal temperature variation, dengue is 
predominantly a  summer disease.
So it is fair to say that these environmental 
fluctuations play a key role in determining beginning and end of epidemic 
periods.
This climatic influence is exerted mainly through its effects on the force of 
infection, which cannot be taken as constant~\cite{reiner_time-varying_2014} 
but rather as a seasonal (oscillating) function of time.
The long term dynamics of dengue is also modulated by the alternation of virus 
types in circulation. Demographics also plays a role in replenishing the 
population of susceptibles. 

The attack ratio (AR) of a disease is a measure of morbidity defined as the 
number of new cases divided by the population at risk.
% LM: Maybe a citation here? It's ridiculous I know, but referees can be a pain.
For dengue epidemics, it can be difficult to calculate the AR due 
to the lack of knowledge of the population at risk.
The population at risk in this case is the number of susceptibles 
to the circulating virus type(s) before a given epidemic. 
Thus, in order to calculate the attack ratio, we need to determine the number 
of susceptibles to the circulating virus types right before the epidemic, which 
is virtually impossible without regular virological surveys.

The attack ratio is also influenced by the reproductive number of the disease 
~\cite{bacaer_final_2009, katriel_attack_2012}, which is closely associated 
with the force of infection.
Thus the incorporation of the effective reproduction number, \rr, as a function 
of time, is crucial to an accurate estimation of the AR of seasonal diseases 
like dengue and Influenza.

Other methods for estimating the  number of susceptibles while 
accommodating time-varying force of infection have been proposed 
before, for measles~\cite{bjornstad_dynamics_2002, 
wallinga_reconstruction_2003}, a disease that shows remarkable seasonality.
These methods try to reconstruct the entire 
series of infectious and susceptibles from case data using deterministic models 
and generally work well for measles because there is a one-to-one 
relationship between exposure and immunity, since measles is caused by a 
single-strain pathogen.
Recently, methods in the same fashion were developed for dengue when 
serotype-specific data is available~\cite{Reiner2014}.
When such data is not available, the series of susceptibles to all possible
serotypes, cannot be reconstructed based solely on a deterministic transmission 
model, since the arrival/re-emergence of  new serotypes, an intrinsically 
stochastic events, can drastically change the pool of susceptibles, throwing off 
any sequential estimation based on the  incidence dynamics.

In this paper, we propose a new approach to estimate the number (fraction) of 
susceptibles using a simplified model of dengue transmission based on a 
single-strain Susceptible-Infectious-Removed (SIR) model with time-varying 
infection rate.
In order to bypass the limitations of not knowing the serotype-specific 
seroprevalence and the exact behaviour of the force of infection through time, 
we
propose to inform the time-varying transmissibility using the \rr series 
derived from the notification data~\cite{nishiura}.
We extend a Bayesian framework previously used to estimate the number 
of susceptibles in Influenza epidemics in Europe~\cite{pone2011} to include 
time-varying force of infection and derive a probability distribution  for 
\rr to accommodate uncertainty in the estimates. 
Then, from the incidence series and the population at risk, we  calculate the 
attack ratio for each epidemic.
We apply our method to estimate $S_0$ before every major dengue epidemic in the 
city of Rio de Janeiro, Brazil in the last 18 years.

\section*{Methods}

In this section we will start by describing the data and then the method used 
to estimate the effective reproductive number, $R_t$, from the data and obtain 
its posterior distribution. 
We then proceed to describe the Susceptible-Infectious-Recovered (SIR) model 
used to represent the aggregated disease incidence and how $R_t$ can be 
integrated into the model to allow for time varying force of infection. 
Next, an approach to approximate the posterior distributions 
of the numbers of susceptible to the main circulating dengue viruses for each 
epidemic is detailed.
Finally, we discuss how to estimate the attack ratio of each 
epidemic using the estimated susceptible fraction and the observed incidence.

\subsection*{Data} 

The data used in this paper consists of time 
series of weekly notified cases of dengue for the 
city of Rio de Janeiro from 1996 to 2014. The cases are notified 
based only on clinical criteria.
Laboratory confirmation and serotype 
information are available only for a very small sample and only on recent 
years (2010-2013).
For the parameter estimation  procedures incidence was normalized by dividing 
the number of cases reported by the total city population at each year as given 
by the census (Census Bureau, Brazilian Institute of Geography and Statistics, 
\url{http://www.ibge.gov.br/english/}).

\subsection*{Estimating the effective reproductive number ($R_t$)}

In monitoring of infectious diseases, it is important to assess whether the 
incidence of a  particular disease is increasing significantly, in order to 
decide to take preventive measures.
The effective reproductive number at time $t$, \rr, can be understood as a 
real-time estimate of the basic reproductive number ($R_{0}$) and is defined as 
the average number of secondary cases per primary case at time $t$.

Let $Y_t$ be the number of reported disease cases for a particular time $t \in 
(0, T)$.
Nishiura el al. (2010)~\cite{nishiura} extend the theory developed by 
Stallybrass et al. (1931)~\cite{stallybrass} and propose to estimate \rr as
\begin{equation}
\label{eq:Rtestimate}
R_t = \left( \frac{Y_{t+1}}{Y_t}\right)^{1/n}
\end{equation}
where $n$ is taken to be the ratio between the length of reporting interval and 
the mean generation time of the disease.
Here we are interested in the simpler case $n=1$.
If \rr is to be used as a decision tool, however, one needs to be able to 
quantify the 
uncertainty about estimate in equation~\ref{eq:Rtestimate}. 
Here we detail how to obtain credibility intervals for \rr under the assumption 
that the counts $Y_t$ are Poisson distributed for all $t$.

We explore the approach of Ederer and Mantel~\cite{mantel}, whose objective is 
to obtain 
confidence intervals for the ratio of two Poisson counts. 
Let $Y_{t} \sim Poisson(\lambda_t)$ and $Y_{t+1} \sim Poisson(\lambda_{t+1})$ 
and define $S = Y_{t} + Y_{t+1}$.
The authors note that by conditioning on the sum $S$
\begin{align}
\label{eq:binlike}
Y_{t+1} | S &\sim Binomial(S, \theta_t) \\
\theta_t &= \frac{\lambda_{t+1}}{\lambda_{t} + \lambda_{t+1}}
\end{align}
Let $c_{\alpha}(\theta_t) = \{\theta_t^{(L)} , \theta_t^{(U)} \}$ be such that 
$Pr(\theta_t^{(L)}<\theta_t <\theta_t^{(U)}) = \alpha$.
Analogously, define $c_{\alpha}(R_t) = \{R_t^{(L)} , R_t^{(U)} \}$ such that 
$Pr(R_t^{(L)}<R_t<R_t^{(U)}) = \alpha$.
Ederer and Mantel (1974)~\cite{mantel} show that one can construct a $100\alpha 
\%$ confidence interval for \rr by noting that
\begin{equation}
\label{eq:confRt}
 R_t^{(L)} = \frac{\theta_t^{(L)}}{(1-\theta_t^{(L)})} \quad \text{and} \quad 
R_t^{(U)} = \frac{\theta_t^{(U)}}{(1-\theta_t^{(U)})}\\
\end{equation}
Because the transform from $\theta$ to \rr is monotonically 
increasing, the result holds for confidence and credibility intervals alike.

Many authors have chosen to quantify the uncertainty about $\theta$ 
following orthodox approaches  (see for example~\cite{wilson} 
and~\cite{clopper}) mainly for simplicity.
We choose instead to take a Bayesian approach and use the  $100\alpha \%$ 
posterior credibility interval for $\theta_t$ as $c_{\alpha}(\theta_t)$.
If we choose the conjugate beta prior with parameters $a_0$ and $b_0$ for the 
binomial likelihood in (\ref{eq:binlike}), the posterior distribution for 
$\theta_t$ is
\begin{equation}
\label{eq:thetapost}
p(\theta_t| Y_{t+1}, S) \sim Beta(Y_{t+1} + a_0, Y_t + b_0)
\end{equation}
Combining equations~(\ref{eq:confRt}) and~(\ref{eq:thetapost}) 
tells us that the induced posterior distribution of $R_t$ is 
a beta prime (or inverted beta) with parameters $ a_1 = Y_{t+1} + a_0$ and $b_1 
=  Y_t + b_0$~\cite{dubey1970}.
The density of the induced distribution is then 
\begin{equation}
\label{eq:densityMantel}
f_P(R_t| a_1, b_1) = \frac{\Gamma(a_1 + b_1)}{\Gamma(a_1)\Gamma(b_1)} R_t^{a_1 
- 
1} (1 + R_t)^{-(a_1 + b_1)}
\end{equation}
Thus, the expectation of \rr is $a_1/(b_1 - 1)$ and its variance is 
$a_1(a_1 + b_1 - 1)/\left((b_1 - 2)(b_1 - 1)^2 \right) $.
Note that this result holds only for $n = 1$.
Sampling from the posterior in~(\ref{eq:densityMantel}) can be made 
straightforward by first sampling from~(\ref{eq:thetapost}) and then applying 
the transform in~(\ref{eq:confRt}).
Also, one can choose $a_0$ and $b_0$ so as to elicit meaningful prior 
distributions for $R_t$.
We show how to elicit the prior for $R_t$ from specified prior mean and variance
or coefficient of variation in the Appendix.

Also, since $R_t > 1$ indicates sustained transmission, one may be 
interested in computing the probability of this event.
This can be easily achieved by integrating~(\ref{eq:densityMantel}) over the 
appropriate interval.
By noting that
\begin{align}
\label{cumprobMantel}
Pr(R_t > 1) &= 1 - \int_0^1 f_P(r)dr \\
            &= 1- Pr(\theta_t < \frac{1}{2})
\end{align}
one can compute the desired probability while avoiding dealing with the density 
in~(\ref{eq:densityMantel}) directly.

\subsection*{Mathematical modelling} % The model

A Susceptible-Infectious-Removed (SIR) model is proposed to model dengue 
dynamics.
In the traditional formulation of the model, transmission is governed by a 
constant transmission rate $\beta$ and recovery happens at a rate $\tau$.

For our analysis we chose to let the force of infection vary with time, just 
as it does in the actual epidemics, as seen in the data. So as the epidemic 
progresses, the effective transmission  rate changes and is 
given by 
\begin{equation} 
 \label{eq:effbeta}
 \beta(t) = \frac{R_t\cdot\tau}{S_0}
\end{equation}
where $R_t$ is the effective reproductive number, estimated as 
in~\ref{eq:Rtestimate}.
The complete model with the time-varying force of infection is given by
the system of ordinary differential equations:
\begin{align}
   \label{eq:model}
 \frac{dS}{dt} &= -\beta(t)SI \\     \nonumber
 \frac{dI}{dt} &= \beta(t)SI - \tau I&\\      \nonumber
 \frac{dR}{dt} &= \tau I&
\end{align}  
where $S + I + R = 1 \: \forall\: t$. % \in [T_0, T_1]$. 
Of course, this is a rather simplified model, in which, for instance, the 
vector is omitted.
The rationale for this simplification is based on the ability of the 
empirically derived $R_t$  to incorporate the effects of the fluctuating vector 
populations.
Also, although there are multiple circulating serotypes, our approach
can not discriminate between them due to the lack of serotype-specific data.
Nevertheless, this modelling strategy can still provide some insight into the 
disease dynamics and allows us to estimate the initial fraction of susceptibles 
$S_0$, a key epidemiological parameter.

\subsection*{Bayesian parameter estimation}

We take a Bayesian approach to the estimation of $S_0$.
First the incidence time series was divided into $J=13$ 
epidemic windows that corresponded to significant raises in incidence and 
normalized to lie on the $[0,1]$ interval.
For a given interval $j = \{ t_j^{\text{start}}, t_j^{\text{end}} \} $ we 
observe an incidence time series $\mathbf{Y_{j}}$.
We are thus interested in the posterior distribution
\begin{equation}
 \label{eq:S0post}
 p(S_{0j}|\mathbf{Y_{j}}) \propto l(\mathbf{Y_{j}}|S_{0j}, R_t, m, \tau 
)\pi(S_{0j}) 
\end{equation}
The likelihood $l(\mathbf{Y_{j}}|\cdot)$ is assumed to be a Normal distribution 
with fixed variance $\sigma^2$.
In this estimation procedure we kept $R_t$ fixed at fixed at the posterior mean 
obtained as described above and fixed $\tau = 1/7\: \text{days}^{-1}$.
To complete the inference, we need to specify prior distributions for the 
parameters of interest.
We place a flat $\text{Beta}(1, 1)$ prior on $S_{0j}\:\forall j$

To approximate the posterior in~(\ref{eq:S0post}) we use Markov chain Monte 
Carlo techniques implemented in the Bayesian inference with Python 
(BIP)~\cite{pone2011} available at  
\url{http://code.google.com/p/bayesian-inference/}.
BIP uses a Differential Evolution Adaptive Metropolis (DREAM)~\cite{vrugt2008} 
scheme that efficiently samples from high-dimensional joint distributions using 
multiple adaptive chains running in parallel with delayed rejection.
Also, as the numerical integration routine implemented within BIP needs 
$\beta(t)$ to be available at 
arbitrary values of $t$, i.e., as continuous function of time because of the 
variable step size, we used linear interpolation to obtain values of $R_t$ for 
any time point.
In this study we used one chain per parameter, i.e, 3 chains for each run.
The chains were run until $5000$ samples were obtained after discarding $500$ 
burn-in samples.
Convergence of the parallel chains was verified at every 100 iterations by the 
calculation of the Gelman-Rubin's R (potential scale 
reduction factor), which approaches $1$ at convergence~\cite{brooks1998}.

\subsection*{Calculating the attack ratio}

The attack ratio of an epidemic is defined by the number of infections divided 
by the size of the population at risk.

\begin{equation}
\label{eq:AR}
A=\frac{\text{\# cases}}{\text{Population at risk}} 
\end{equation}

Based on what has been discussed so far, we can rewrite (\ref{eq:AR}) for 
each epidemic $j$ as
\begin{equation}
\label{eq:AR2}
 A_{j}=\frac{\sum Y_j}{S_{0j}}
\end{equation}
where $S_{0j}$ is the number of susceptibles before each epidemic $j$, which we
estimated before.

Python and R code to perform all the analyses described above is publicly 
available at~\url{https://github.com/fccoelho/paperLM1}.

\section*{Results and Discussion}

In this paper we propose a method to bypass the lack of serotype-specific case 
data by informing the time-varying force of infection with the instantaneous 
reproductive number, \rr which we calculate from aggregated data.
The main contribution of this paper can be summarized in the following items: 
(i) we show a method to quantify uncertainty 
about \rr that is readily applicable to other diseases and; (ii) we propose to 
use \rr to inform a dynamic epidemic model with time-varying force of infection 
in order to gain insight into the attack ratio of each epidemic; (iii) We 
propose an estimation procedure for circulating serotype's $S_0$ from 
aggregate case data, which is robust to epidemic sizes; We estimate the AR 
for 18 years of Dengue epidemics. 

\begin{center}
\begin{figure}[!h]
 \centering
 \includegraphics[width=16cm]{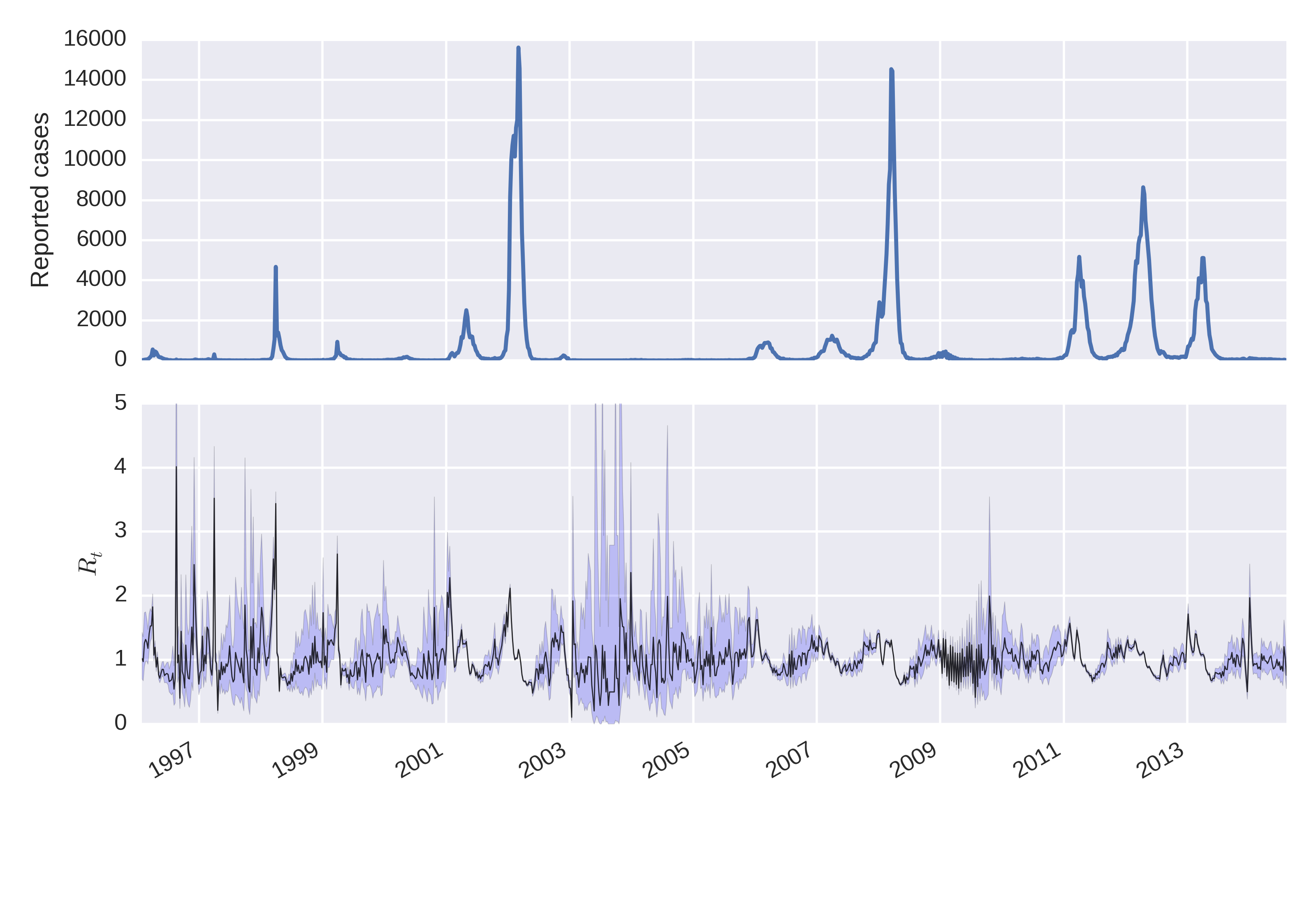}
 \caption{{\bf Estimated time-series for $R_t$, along with 95\% credible 
intervals.} Top panel show reported cases from which $R_t$ is estimated.
As expected, uncertainty about \rr is greater when the case counts are low, for 
instance in the period $2003$--$2006$, which represented a big hiatus between 
major epidemics.
The intrinsic variability of \rr can be used to inform the time-varying force 
of infection, since it reflects variation in the vector population and other 
environmental factors such as temperature and seasonal variation.}
\label{fig:rtseries}
\end{figure}
\end{center}

Figure~\ref{fig:rtseries}  shows the $R_t$ series, according to 
(\ref{eq:Rtestimate})~\cite{nishiura} along with the confidence bands derived 
in this paper. 
It can be seen that the inter-epidemic periods are characterized by $R_t$ being 
indistinguishable from 1.
Due to the intrinsic variability of the \rr series, the examination of its 
credible intervals is essential to identify periods of sustained transmission.
The wider intervals between epidemics are due to the scarcity of cases during 
these periods.
The method to quantify uncertainty proposed here provides more conservative 
credibility intervals, and therefore offers protection against false alarms.

A key epidemiological quantity is the attack ratio (AR) of an epidemic, a 
measure of morbidity and speed of spread which can be used to predict epidemic 
size and help efficient Public Health planning.
The AR depends fundamentally on the population at risk, which in the case of 
dengue is every naive (to a particular serotype) individual in the population.
Estimating the initial susceptible fraction $S_0$ for each epidemic is thus 
central to the estimation of the AR.
Methods for estimating the  number of susceptibles have been proposed 
before, for other diseases~\cite{bjornstad_dynamics_2002, 
wallinga_reconstruction_2003}.
These methods try to reconstruct the entire 
series of infectious and susceptibles for measles 
outbreaks from case data.
In the case of dengue, the full (multi-year/multi-epidemic)
series of susceptibles to all possible serotypes, cannot be reconstructed based 
solely on a deterministic transmission model, since the arrival/re-emergence of 
new serotypes (which are a stochastic events) can change drastically the pool 
of susceptibles throwing off any sequential estimation based on the incidence
dynamics.

Since there is very limited information regarding the actual proportions of 
each virus in circulation and most information available is about the 
predominant serotypes for some epidemics in the period of 
study only~\cite{macedo_virological_2013}, we propose the use of a 
simplified a single strain model.
The main argument we put forward is that by conditioning on the \rr series, we 
implicitly take into account the variability introduced by the 
co-circulation of multiple serotypes and heterogeneous levels of immunity in 
the general population.
We sought to deal with all important sources of uncertainty impinging on the 
estimation of the AR of a dengue epidemic, but not all could be satisfactorily 
addressed in this analysis.
For instance, in any given epidemic there is a large number of 
mild and asymptomatic cases, which nevertheless acquire immunity.
It is estimated that for every case reported, up to 10-20 are not seen by health 
authorities~\cite{luz_disability_2009}.
Another source of uncertainty is under-reporting of diagnosed cases, which is a 
serious issue in the health care systems of many developing countries such as 
Brazil. 
Duarte and Fran\c{c}a (2006)~\cite{duarte_data_2006}, estimated 
the sensitivity of Dengue reporting for hospitalized patients in 
Belo-Horizonte, Brazil to be of 63\%, meaning that approximately 37\% of the 
suspected Dengue cases go unreported.  
Lastly, demography and migrations affect the number of susceptible in ways 
which are not easy to fully determine.

\begin{center}
\begin{figure}
 \subfigure[2010]{\includegraphics[width=8cm, 
height=4.82cm]{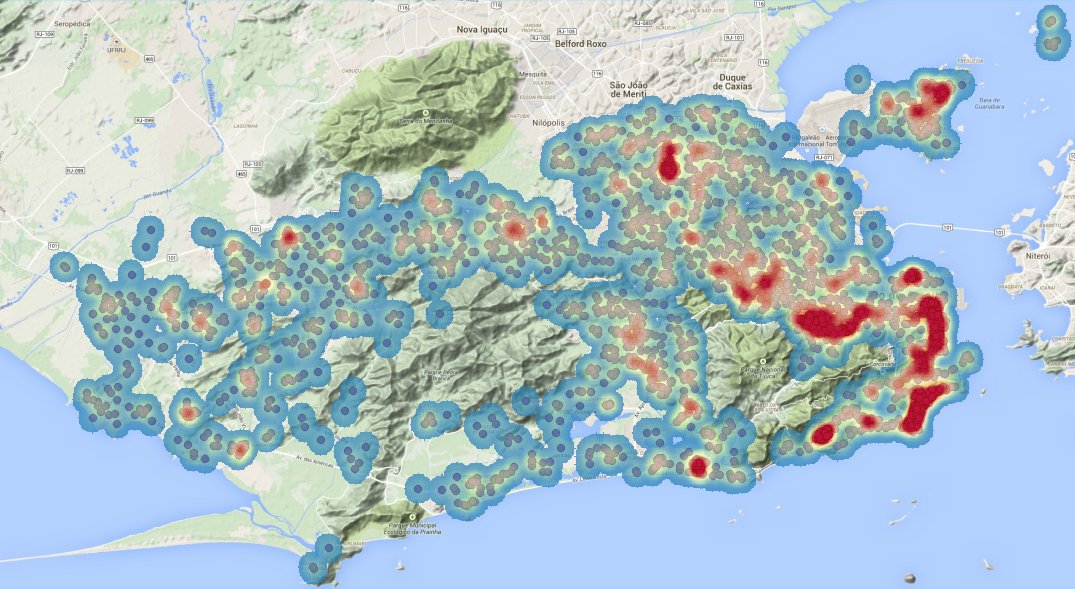}
% heatmap2010.jpg: 1075x589 pixel, 96dpi, 28.44x15.58 cm, bb=0 0 806 442
}
  \subfigure[2011]{\includegraphics[width=8cm]{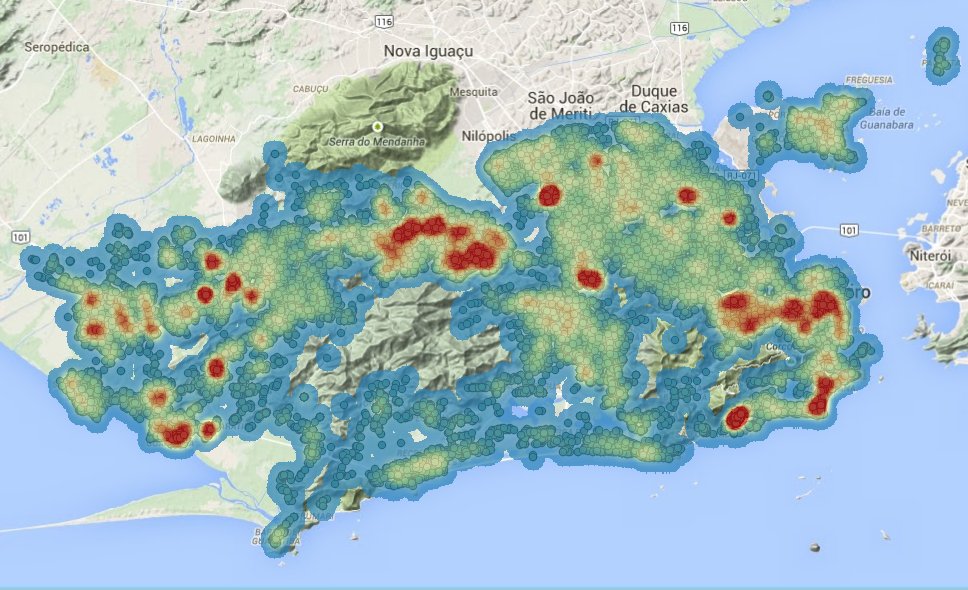}}
  \subfigure[2012]{\includegraphics[width=8cm]{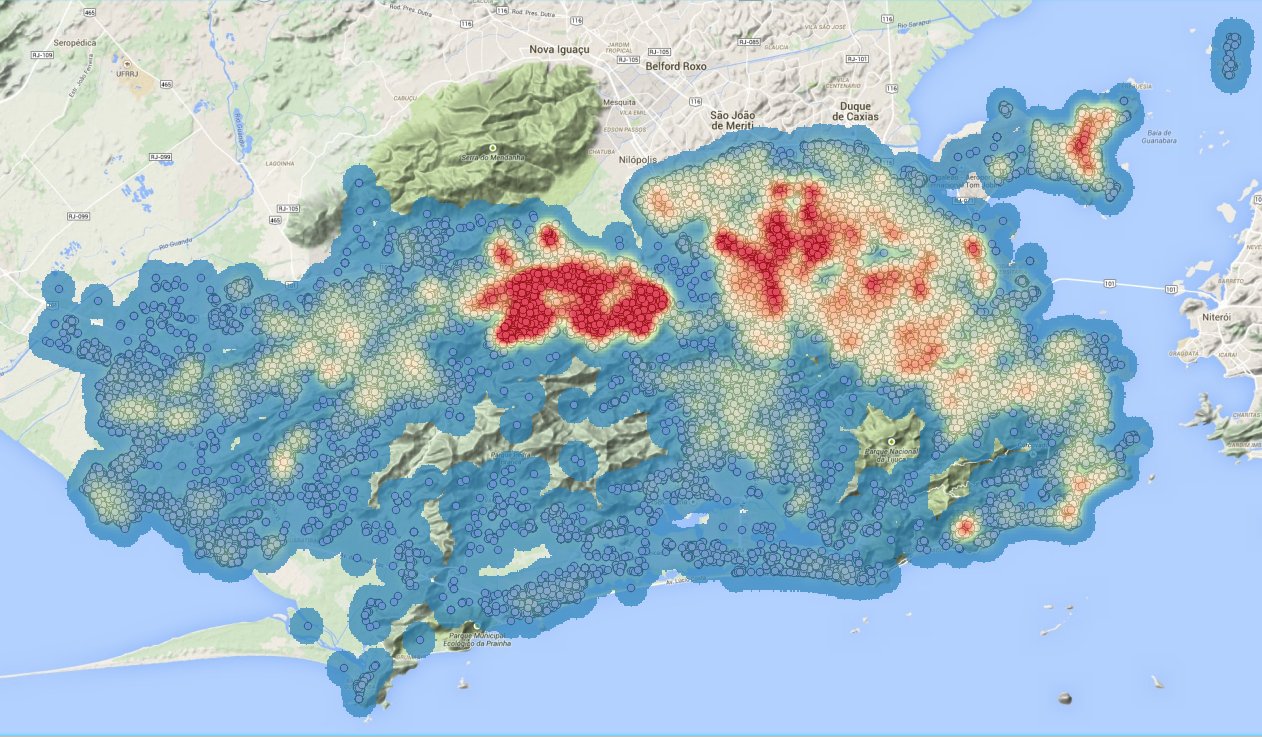}
% heatmap2010.jpg: 1075x589 pixel, 96dpi, 28.44x15.58 cm, bb=0 0 806 442
}
  \subfigure[2013]{\includegraphics[width=8cm]{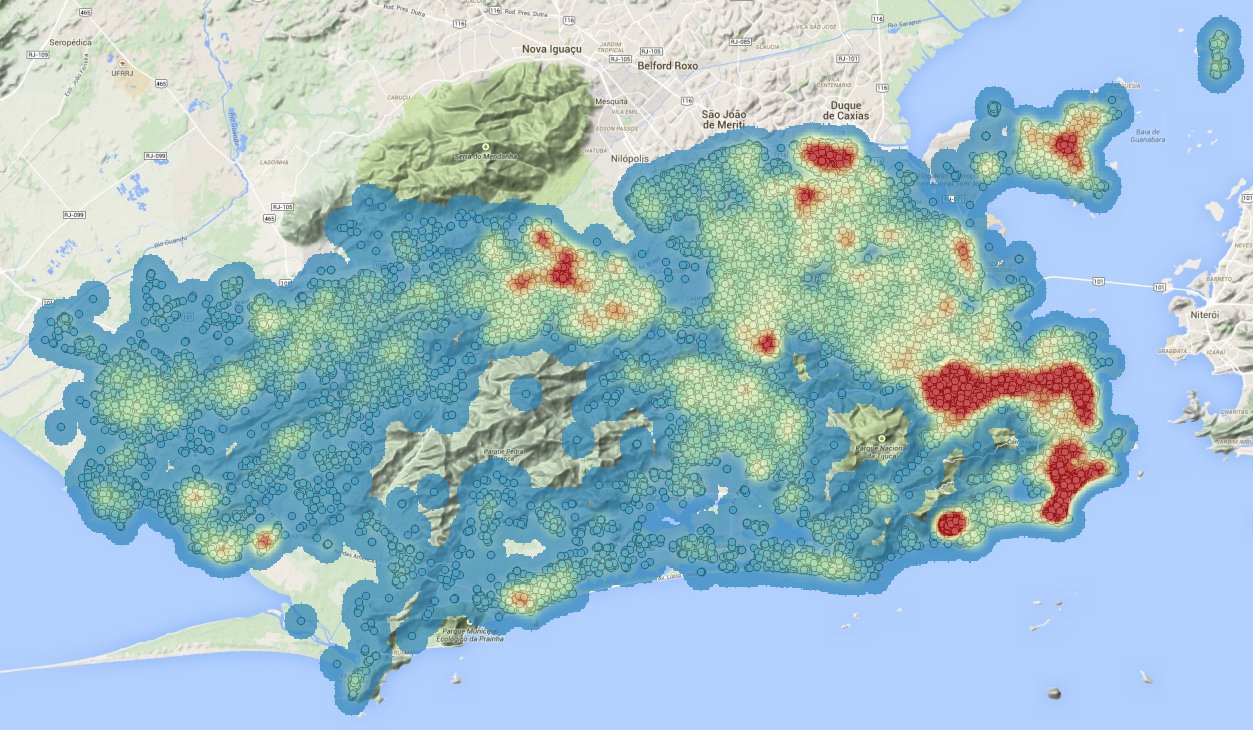}}
  \caption{{\bf Maps showing the incidence of dengue in the city of Rio de 
Janeiro from 2010 to 2013.} Circles indicate individual notified cases. A 
heatmap is overlayed on the maps showing absolute density of cases.
It can be seen that several areas of the city were affected and no region seems 
to be free of transmission risk.
This suggests that although transmission risk varies spatially, there is 
significant exposure over the entire city.}
  \label{fig:mapas}
\end{figure}
\end{center}

\begin{figure}[!ht]
\begin{center}
\includegraphics[width=\textwidth]{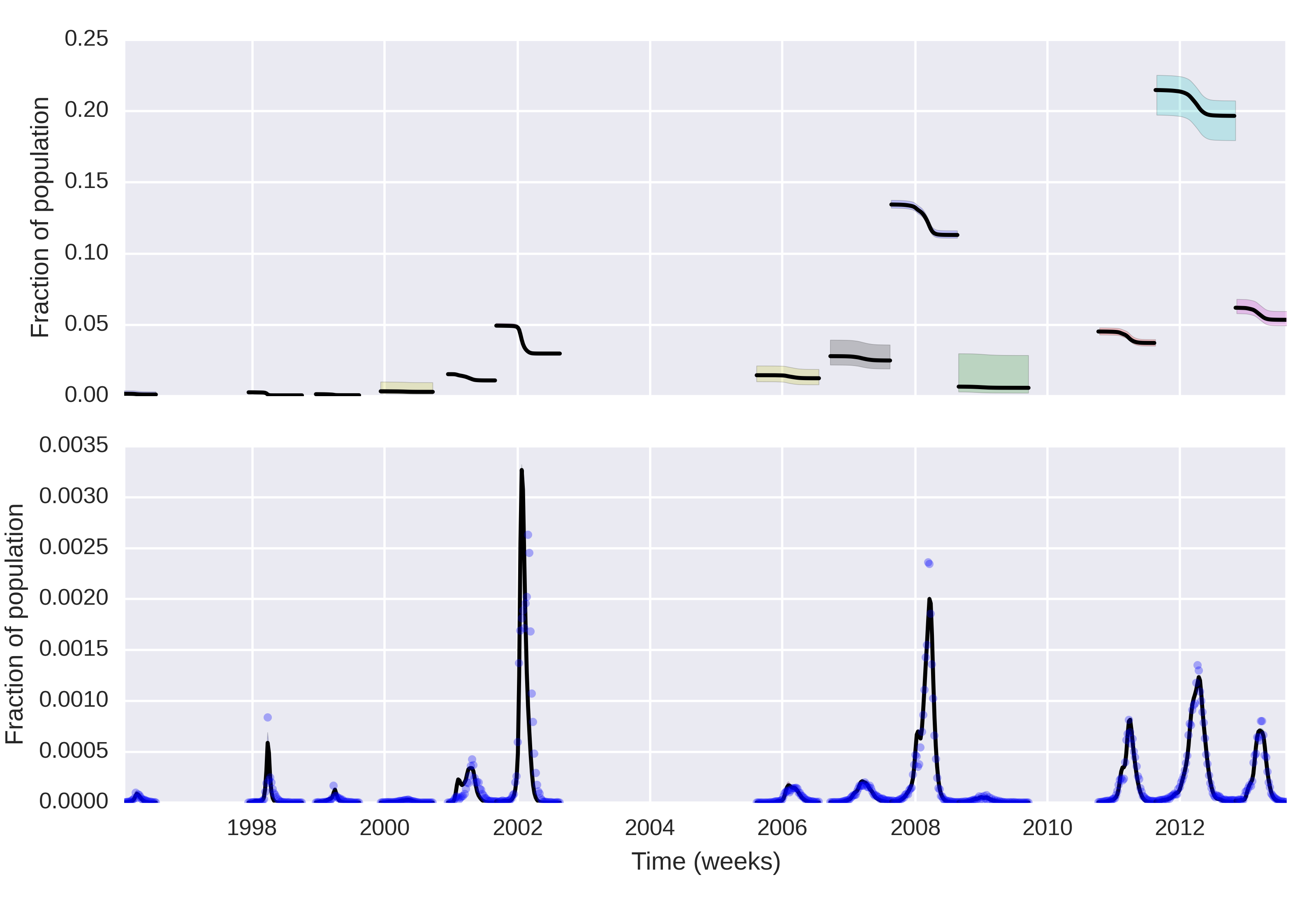}
\end{center}
\caption{
{\bf Susceptibles and Infectious posterior curves.}  The curves were estimated 
only for the periods where $R_t> 1$.  The susceptible curves in the top panel 
reflect the prevalence of fraction of susceptibles to circulating strain(s) for 
each epidemic/outbreak.
In the lower panel, we see the posterior distribution 
of infectious curves, represented by its median and 95\% credible interval.
Credible intervals are very narrow, and can be hard to distinguish from the 
median line. 
Dots show the observed cases, scaled as fractions of the entire 
population.
}
\label{Fig:S0}
\end{figure}

Figure~\ref{Fig:S0}, shows the model from (\ref{eq:model}) fitted to the data.
Despite its limitations, our simplified model fits the data well.
In it we can see that the susceptibles series in each epidemic starts at the 
estimated level of $S_0$.
The proportion of susceptibles may seem low, but we 
must remember that these estimates are being affected by an unknown 
under-reporting factor, which experts suggest is somewhere between 5 and 10, 
i.e. for every case observed there are 5 or 10 unobserved.
Since this under-reporting affects both the numerator and denominator of 
(\ref{eq:AR2}), its effects should cancel out, giving us an unbiased attack 
ratio estimate.
One other possible source of bias which would lead to the underestimation of 
$S_0$ could come from a significant part of the population not being exposed to 
the disease.
However, as we can see in Figure~\ref{fig:mapas}, despite the differences in 
intensity (incidence), the entire city seems to be at risk, with no 
particularly ``protected'' areas, at least in the last four epidemics. 

Table~\ref{tab:AR} contains the attack ratios and medians of the $S_0$ 
estimated for each epidemic/outbreak.
It is interesting to notice that the larger epidemics, in terms of peak size are 
not the one with the greater attack ratios.
This stresses the importance of knowing the immunological structure of the 
population. Knowing the $S_0$ for the circulating viruses we can order to more 
accurately assess the potential impact of a coming epidemic, since particularly 
virulent types, can be rendered less of a threat by a low $S_0$.

We hope that the results presented in paper will motivate public 
health authorities to invest in annual serological surveys, to determine the 
susceptibility profile to each dengue virus as well as to estimate the 
under-reporting factor of the notification system.

\begin{table}[!ht]
\caption{
{\bf Median attack ratio and 95\% credibility intervals calculated according to 
(\ref{eq:AR2})}. 
Values are presented as percentage of total population. 
$^\dag$: Year corresponds to the start of the epidemic, however the peak of cases
may occur in the following year.
$^\ddag$: Susceptible fraction.
These results show considerable variation in AR between epidemics, consistent with
the accquiring and loss of serotype-specific immunity.}
\begin{center}
\begin{tabular}{c|c|c}
\hline
Year$^\dag$ & median Attack Ratio & $S_0^\ddag$ \\
\hline
1996 & 0.39 (0.17-0.54) & 0.00171(0.0012-0.0038)\\
1997 & 0.87 (0.74-0.87) & 0.00273(0.0027-0.0032)\\
1998 & 0.5 (0.49-0.5) & 0.00142(0.0014-0.0014)\\
1999 & 0.11 (0.037-0.2) & 0.00345(0.0018-0.01)\\
2000 & 0.25 (0.24-0.27) & 0.0155(0.015-0.016)\\
2001 & 0.48 (0.47-0.49) & 0.0495(0.048-0.051)\\
2005 & 0.15 (0.1-0.21) & 0.0147(0.01-0.021)\\
2006 & 0.11 (0.08-0.14) & 0.0281(0.022-0.037)\\
2007 & 0.15 (0.15-0.15) & 0.135(0.13-0.14)\\
2008 & 0.14 (0.031-0.31) & 0.00672(0.003-0.024)\\
2010 & 0.18 (0.17-0.19) & 0.0454(0.043-0.048)\\
2011 & 0.086 (0.082-0.094) & 0.215(0.2-0.23)\\
2012 & 0.14 (0.13-0.15) & 0.0621(0.058-0.068)\\
\hline
\end{tabular}
\end{center}
\label{tab:AR}
\end{table}

\newpage
\section*{Acknowledgements}
LMC is grateful to Dr. Leonardo Bastos for useful discussions on the 
posterior inference for \rr.
The authors are also grateful to Claudia T. 
Code\c{c}o for helpful discussions about the manuscript.

\newpage
\bibliography{lm1}

\section*{Appendix}
\subsection*{A remark on prior distributions and tail behaviour of the 
distribution of $R_t$}
\label{sec:tails}
There are a number of approaches to deriving the distribution of \rr.
Alternatively to the approach described in the main text~\cite{mantel}, one 
could use the conditional distribution of \rr on 
$Y_{t+1}$ and $Y_t$ as defined in equation A7 of Nishiura et 
al.~\cite{nishiura}:
\begin{equation}
\label{seq:unorm}
f_{R}(R_{t}) = (Y_tR_{t})^{Y_{t+1}} e^{-Y_tR_{t}}
\end{equation}
Noticing the kernel of (\ref{seq:unorm}) is that of a gamma distribution with 
$a_2 = Y_{t+1}+1$ and $b_2 = Y_t$, we obtain a proper density from which to 
construct $c_{\alpha}(R_t)$, simply by computing the appropriate quantiles of 
said distribution.
 This density is
\begin{equation}
\label{seq:densityNishiura}
f_N(R_t| a_2, b_2) =  \frac{b_2^{a_2}}{\Gamma(a_2)} R_t^{a_2-1} e^{-b_2 R_t}
\end{equation}

In order to decide which approach to take, it may be of use analysing the 
tail behaviour of the derived distributions for \rr. 
Consider the case of using a flat $Uniform(0, 1)$ prior for $\theta_t$.
With $a_0 = b_0 = 1$, $a_1 = a_2$ and $b_1 = b_2 + 1$.
The beta prime (inverse beta distribution) will have heavier tails compared to 
the conditional distribution proposed by~\cite{nishiura}, thus providing more 
conservative confidence/credibility intervals.
To see that one needs simply take the ratio of the Beta prime and Gamma 
(unnormalized) densities and evaluate the limit as $R_t$ goes to infinity:
\begin{equation}
 \label{seq:densityratio}
 \lim_{R_t\to\infty}\frac{f_P(R_t| a_1, b_1)}{f_N(R_t| a_2, b_2)} =  
\lim_{R_t\to\infty}\frac{e^{Y_{t}R_t}}{(1 +R_t)^{Y_{t} + Y_{t +1}+2}} = \infty
\end{equation}
Finally, note that we deliberately construct $c_{\alpha}(R_{t})$ as a 
equal-tailed $100\alpha\%$ credible set, rather than a less conservative 
highest posterior density (HPD) interval.

As a side note, the Bayesian approach presented in this 
paper will give similar results to orthodox confidence intervals~\cite{wilson} 
and~\cite{clopper} for $Y_{t+1}$ and $Y_t >> 1$.
Under the flat  uniform prior for $\theta_t$, the Bayesian posterior 
credibility 
interval is nearly indistinguishable from the confidence interval proposed by 
Clopper \& Pearson (1931)~\cite{clopper} for $Y_{t+1}, Y_t > 20$.
Note also that the uniform prior ($Beta(1, 1)$) for $\theta_t$ constitutes a 
poor 
prior 
choice mainly because the induced distribution for \rr is only well-defined for 
$b_0 > 2$.

An advantage of the Bayesian approach is that one can devise prior 
distributions for $\theta_t$ taking advantage of the intuitive parametrization 
and flexibility of the beta family of distributions.
Prior elicitation can also be done for \rr and the hyper-parameters directly 
plugged into the prior for $\theta_t$. 
One can, for example, choose prior mean and variance for \rr and find $a_0$ 
and $b_0$ that satisfy those conditions.
Let $m_0$ and $v_0$ be the prior expectation and variance for $R_t$. 
After some tedious algebra one finds
\begin{align}
\label{seq:elicitation}
a_0 &= \frac{m_0v_0 + m_0^3 + m_0^2}{v_0} \\
b_0 &= \frac{2v_0 + m_0^2 + m_0}{v_0}
\end{align}
If one wants only to specify $m_0$ and the coefficient of variation $c = 
\sqrt{v_0}/ m_0$ for $R_t$ \textit{a priori}, some less boring 
algebra gives:
\begin{align}
\label{seq:elicitationcv}
a_0 &= \frac{m_0^3c^2 + m_0^3 + m_0^2}{m_0^2c^2} \\
b_0 &= \frac{2m_0^2c^2 + m^2 + m}{m_0^2c^2}
\end{align}

This approach thus makes it possible to incorporate epidemiological knowledge 
about disease Biology (e.g. the magnitude of $R_0$) into the computation of \rr.
This may prove particularly important when disease counts are low and/or close 
to the detection threshold.
We provide an R script to perform the above elicitation at 
\url{https://github.com/fccoelho/paperLM1/blob/master/R/elicit_Rt_prior.R}.

\end{document}